\documentclass [12pt]{article}
\usepackage{graphicx}
\usepackage{setspace}
\usepackage{amsmath}
\usepackage[dvipsnames]{xcolor}
\usepackage{float}
\usepackage{enumitem}
\usepackage{subcaption}
\usepackage{amssymb}
\usepackage{hyperref}
\hypersetup{
    colorlinks=true,
    linkcolor=blue,
    filecolor=magenta,      
    urlcolor=cyan}
\usepackage[all]{hypcap}
\usepackage[maxnames=4]{biblatex}
\DeclareFieldFormat[article, book, misc]{title}{\mkbibquote{#1\isdot}}
\newcommand{\abs}[1]{\left\lvert#1\right\rvert}

\renewbibmacro{in:}{}

\title{Pickleball Flight Dynamics}
\author{Kye Emond, Weiran Sun and Tim B.\ Swartz
\thanks{
K.\ Emond is an undergraduate student, Department of Physics,
W.\ Sun is Professor, Department of
Mathematics and T.\ Swartz is Professor,
Department of Statistics and Actuarial
Science, Simon Fraser University, 8888 University Drive, Burnaby
BC, Canada V5A1S6.
All authors have been partially supported by the
Natural Sciences and Engineering Research Council of Canada.
}}

\date{ }
\begin{document}
\maketitle
	
\begin{abstract}
\noindent
This paper considers the flight dynamics of
the ball in the sport of pickleball.
Various simplifications are introduced according to
the features of the game. These simplifications and
some approximations enable straightforward coding to study
aspects of the game such as the trajectory of the ball and
its velocity.
In turn, strategic questions may be
addressed that have not been previously considered.
In particular, our primary research question involves
the preference between playing with the wind versus
against the wind. It is demonstrated that 
playing against the wind is often preferable than playing with the wind.
\end{abstract}
	
\vspace{3mm}

\newpage
\section{INTRODUCTION}
\label{sec:intro}

Pickleball is a relatively new sport. It was invented
in 1985, and in recent years its popularity has taken off.
Pickleball was the fastest growing sport from 2022
to 2023 in the United
States with over 8.9 million participants [1]. 
According to a 2023 report from the Association of Pickleball
Players (APP), nearly 50 million Americans have played pickleball
at least once in the previous year (\url{https://www.theapp.global}).
The game is popular across wide age cohorts at the recreational
level. Pickleball also has
various professional leagues and tours including
Major League Pickleball (MLP).

Despite the popularity of the sport, there has been little
quantitative research on pickleball. [5] consider the impact of strong and weak links 
on success in doubles pickleball.
It is the intention of this paper to add to the sparse literature
with a specific aim of gaining a better understanding
of pickleball flight dynamics. 
[2] consider problems in sports analytics
across major sports.

The topic of projectile motion has a long and
well-studied history [6]. 
The details are complex, especially when considerations are
given to the impact of
air resistance and wind. Projectile motion
models typically involve special functions and differential
equations.
Such work is important
to serious investigations such as ballistics. 
In sport, [3] considers issues of approximate projectile motion in the sports of golf, tennis and badminton. 
However, there does not seem to be any literature on pickleball
flight dynamics; this paper attempts to provide some initial insights on this topic.

In the problem considered here, we take features of the sport
of pickleball into account. This, together with additional
assumptions simplifies our projectile motion model. The
final model is straightforward to code so that various investigations
involving pickleball may be undertaken.
In particular, we look at the impact of the wind in pickleball. Pickleball
is often played outdoors where the choice of ends, and understanding
how to play
in the wind become issues of strategy. 
Our primary research question involves
the preference between playing with the wind versus
playing against the wind where it is demonstrated that playing against
the wind is preferable in many contexts.
This problem in pickleball strategy
does not seem to have been previously addressed.

In Section~\ref{sec:Formulation}, we provide a description of the relevant details of
the pickleball court, and features of interest.
We also define the relevant input variables to the projectile motion model.
In Section~\ref{sec:motionmodel}, 
the basics of the pickleball motion model are described. In particular,
we explain how features and strategies in the sport allow us to calculate 
input variables that are not immediately available.
In Section~\ref{sec:Applications}, we look at various pickleball applications. 
In particular, we investigate pickleball trajectory and
pickleball velocity under various conditions. 
We then discuss a question of
strategy in terms of whether it is better to play against the wind
or with the wind. The work indicates that a strategic advantage
is often conferred when playing against the wind.
We conclude with a short discussion in Section~\ref{sec:Discussion}. Details regarding modelling and simulations are left to the Appendix.

\section{PROBLEM FORMULATION}
\label{sec:Formulation}

\begin{figure}[ht]
\centering
\vspace{-30mm}
\includegraphics[scale = 0.62]{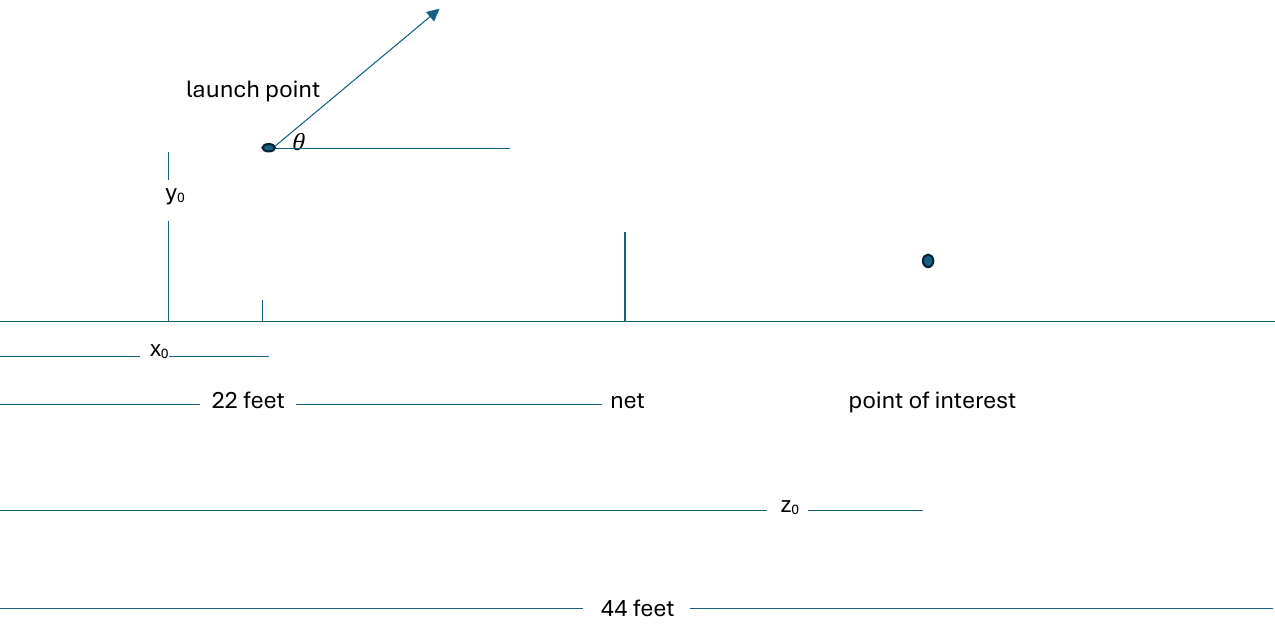}
\vspace{-55mm}
\caption{Configuration of the court and the variables related
to the flight of the pickleball.}
\label{fig:Setup}
\end{figure}

Figure~\ref{fig:Setup} provides the relevant details of the pickleball court
and features of interest.
The pickleball court is 44 feet long which is divided into two equal
halves by a net. The net is 3 feet tall at the ends although this
detail is not important for our motion model.

In Figure~\ref{fig:Setup}, a launch point is depicted on the left side of the
court. This is the location from which the
player of interest strikes the
pickleball. The location is marked $x_0$ feet from the left endline and serves
as an input variable for our investigation. We have the
constraint $x_0 \in (0, 15)$ feet where we note that 
the 15 foot mark denotes the 
beginning of the non-volley zone (i.e.\ the closest
point to the net that the player should approach).
The player strikes the ball at height $y_0$. We consider
$y_0 \in (1,3)$ feet as a range for the height at
which the pickleball is struck. Although the pickleball can be struck
from higher heights, this range corresponds to the situation where
the ball is hit in an upwards trajectory. 
Further, the ball is struck at launch angle $\theta$.
For our purposes, we consider $\theta \in (10^\circ,30^{\circ})$.
An angle larger than $30^{\circ}$ either represents a lob
shot or a mishit, two shots that are not relevant
to this investigation. 

In Figure~\ref{fig:Setup}, we also depict the opponent (i.e.\ the point of
interest) on the right side
of the court whose horizontal position is given by $z_0$ feet
from the left endline. Later,
we are interested in the opponent's ability to
hit the struck ball.
Since the opponent is not advised to stand in his
non-volley zone, we have the constraint $z_0 \in (29,44)$.

There are two quantities that are relevant to our investigation 
that are not depicted in Figure~\ref{fig:Setup}. First, the wind is a characteristic of 
interest. We make the assumption that the wind blows in a strictly horizontal
direction. Our personal experiences in pickleball suggest that 
playing in winds which are less than 10 mph is largely inconsequential. On the other
hand, playing in wind speeds exceeding 20 mph is extreme and is a situation
that many players avoid. Therefore, we are interested in wind velocities $w$
(i.e.\ speed and direction) in the intermediate
intervals $(-20,-10)$ mph and $(10,20)$ mph.

Second, we require the initial velocity $v_0$ which is velocity that the pickleball is struck
at the launch point. In the related sport of tennis, the average serve
in men's professional tennis (e.g.\ the ATP tour) is estimated at 120 mph.
Unlike tennis, the pickleball paddle is rigid (without strings), and
the ball is hard and compresses only negligibly. Therefore,
the fastest pickleball shots reach instantaneous speeds of roughly 60 mph.

Therefore, to summarize, the input variables that are relevant to 
pickleball flight dynamics
are $(x_0, y_0, \theta, z_0, w, v_0)$.

\newpage
\section{PICKLEBALL MOTION MODEL}
\label{sec:motionmodel}

This section describes the basics of the pickleball motion model.
More details including the associated physics of the model
are provided in the Appendix.

For this investigation, it is convenient
to express the location and the speed of the pickleball in both the $x$ and $y$ coordinates.
We denote the location and speed of the
pickleball by
$x$ and $x'$ in the horizontal direction, and by
$y$ and $y'$ in the vertical direction.

Referring back to the discussion and the
notation in Section~\ref{sec:intro}, the coordinate speeds
are expressed more fully as
\begin{eqnarray}
x'(t, \theta, w, v_0) 
\quad \text{and} \quad
y'(t, \theta, w, v_0).
\label{eq:x'}
\end{eqnarray}


The arguments of the speeds in ~\eqref{eq:x'} 
have common
terms, namely the time from launch $t$, the launch angle $\theta$, 
the wind velocity $w$ and the initial velocity $v_0$.
Of course, and as described in the Appendix, the functions in~\eqref{eq:x'} 
also depend on the features of the pickleball
(e.g.\ weight, size and surface) which determine the
impact of air resistance. Also, the force of gravity comes into
play in the vertical speed but not in the horizontal speed.
In our model, we ignore the impact of spin.

In~\eqref{eq:x'} 
we note that the speed functions depend
on the launch angle $\theta$ and the initial velocity $v_0$.
Since the initial
coordinate speeds only depend on $\theta$ and $v_0$
through the initial coordinate speeds, using trigonometry
in~\eqref{eq:x'},
we may replace $\theta$ and $v_0$ in $x'$ by $v_0 \cos \theta$, and
we may replace $\theta$ and $v_0$ in $y'$ by $v_0 \sin \theta$.
However, we retain the excessive notation in~\eqref{eq:x'} 
which is helpful
in future considerations.

For the coordinate locations, these are expressed more fully as
\begin{eqnarray}
x(x_0, t, \theta, w, v_0)
\quad \text{and} \quad
y(y_0, t, \theta, w, v_0).
\label{eq:x}
\end{eqnarray}
The functions in~\eqref{eq:x} 
have the same arguments as in (\ref{eq:x'}) 
except that the initial locations $x_0$ and $y_0$ also influence
location at time $t$. 

It may be noted that the relationship between location and velocity
allows us to express the locations functions as
$x(x_0,t,\theta, w, v_0)
= x_0 + \int_{0}^t x'(s, \theta, w, v_0) \ ds$ and
$y(y_0,t,\theta, w, v_0)
= y_0 + \int_{0}^t y'(s, \theta, w, v_0) \ ds$. However,
these expressions do not assist our development since the
integrands are intractable functions.

\subsection{A Pickleball Simplification}
\label{subsec:simple}

A primary interest in our research concerns the issue of playing
in the wind; should you prefer to play with the wind or play against the wind?

Of course, in pickleball, there are various types of shots and these
include lobs, dink shots, smashes, drops, drives, etc. For
the time being, we are going to restrict our attention to drive shots.

With respect to drive shots, we simplify aspects of the motion model
by considering some standard pickleball strategy.
Referring to Figure~\ref{fig:Setup},
we assume that the player on the left hand side 
of the court (i.e.\ the launch point) hits
the ball as hard as possible 
such that the ball would
remain in bounds if left untouched.
This assumption is sensible for drive shots in pickleball.
Players hit the ball hard because high speed shots pose difficulty
for the opponent; in particular, the opponent has less time to react.
Hitting the ball as described, means that the ball, if left untouched,
would land
on the endline on the right hand side of the court.
Therefore, hitting the ball in this manner may be considered optimal
for drive shots in pickleball.

We denote $t_b$ as the hypothetical time that
it would take
the hard hit ball to {\it bounce} on the right endline.
Because the length of the court is 44 feet, we can express this
constraint~as
\begin{eqnarray}
y(y_0, t_b, \theta, w, v_0) = 0,
\qquad
x(x_0, t_b, \theta, w, v_0) = 44.
\label{eq:simple1}
\end{eqnarray}

With equations~
\eqref{eq:simple1}, we are going to investigate various cases
involving the input settings $x_0$, $\theta$ and $w$. In other
words, $x_0$, $\theta$ and $w$ are values that are determined in advance.
Therefore, 
\eqref{eq:simple1} represents two equations
in two unknowns, $t_b$ and $v_0$. Using the model described in
the Appendix and the associated numerical methods, we are
able to solve for
$t_b$ and $v_0$. This is particularly helpful since these
are two quantities for which little is known apriori.

Having solved for $v_0$, we can then consider the equation
\begin{eqnarray}
x(x_0, t, \theta, w, v_0) = z_0
\label{eq:x=z}
\end{eqnarray}
for an unknown time $t$. Equation~(\ref{eq:x=z}) addresses the time that
it takes the ball from when it is struck
to reach the opponent (i.e.\ the location
of interest in Figure~\ref{fig:Setup} which is $z_0$ feet from the left endline).

From (\ref{eq:x=z}), we are able to solve for $t$. 
When $t$ is small, this means that there is little time for the opponent to
react with their return shot. Therefore, the shot would be a very good shot.
Consequently, for wind speeds
$w$ and $-w$, we can assess whether it is better to play
with or against the wind in the context of a drive.
This problem is studied in Section~\ref{subsec:wind}.

\section{APPLICATIONS}
\label{sec:Applications}

\subsection{Pickleball Trajectory}
\label{subsec:traj}

Using the motion model described in the Appendix for drive shots and the 
associated numerical methods,
we are able to compute both the horizontal location
$x(x_0, t, \theta, w, v_0)$
and the vertical location
$y(y_0, t, \theta, w, v_0)$
given the input variables.
The resulting $(x,y)$ coordinates taken over
a sequence of times $t$ allow us to produce trajectory plots.
Note that our code allows us to do this over any set of input variables.

\begin{figure}[ht]
    \centering
    \includegraphics[width=\textwidth]{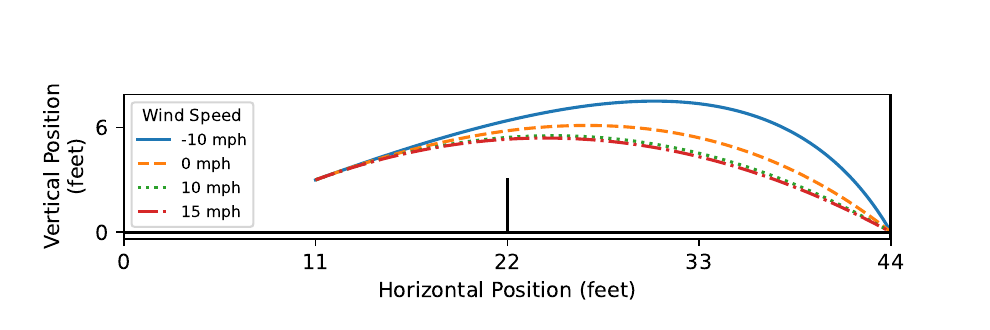}
    \vspace{-10mm}
    \caption{Trajectory of an optimally hit drive shot in four wind
    conditions $w=-10$ mph, $w=0$ mph, $w=10$ mph and $w=15$ mph.
    Other input values are set at
    $x_0 = 11$ feet, $y_0=3$ feet and $\theta = 20$ degrees.}
    \label{fig:trajectory}
\end{figure}

In Figure~\ref{fig:trajectory}, we provide plots for input values
$x_0 = 11$ feet (which corresponds to the middle of the left court),
$y_0 = 3$ feet (which is a typical height from where the ball is hit)
and
$\theta = 20$ degrees (which is a typical launch angle).
Four plots are provided; for wind speeds $w=-10$ mph, $w=0$ mph (no wind), 
$w=10$ mph and $w=15$ mph.
The initial velocity input $v_0$ is evaluated according to the optimality
conditions~\eqref{eq:simple1} described in Section~\ref{subsec:simple}.

In Figure~\ref{fig:trajectory}, we observe that
the trajectories for wind speeds $w=0,10,15$ mph
do not differ greatly. However, when playing against the wind
(i.e.\ $w=-10$ mph), the pickleball flight has greater curvature
with a higher arc. It appears that the pickleball (which is light)
gets held up by the wind. Towards the end of the path
when playing against the wind, the pickleball
is moving more in a downward vertical direction than horizontally.

\subsection{Pickleball Velocity}

\begin{figure}[ht]
    \centering
    \includegraphics[width=\textwidth]{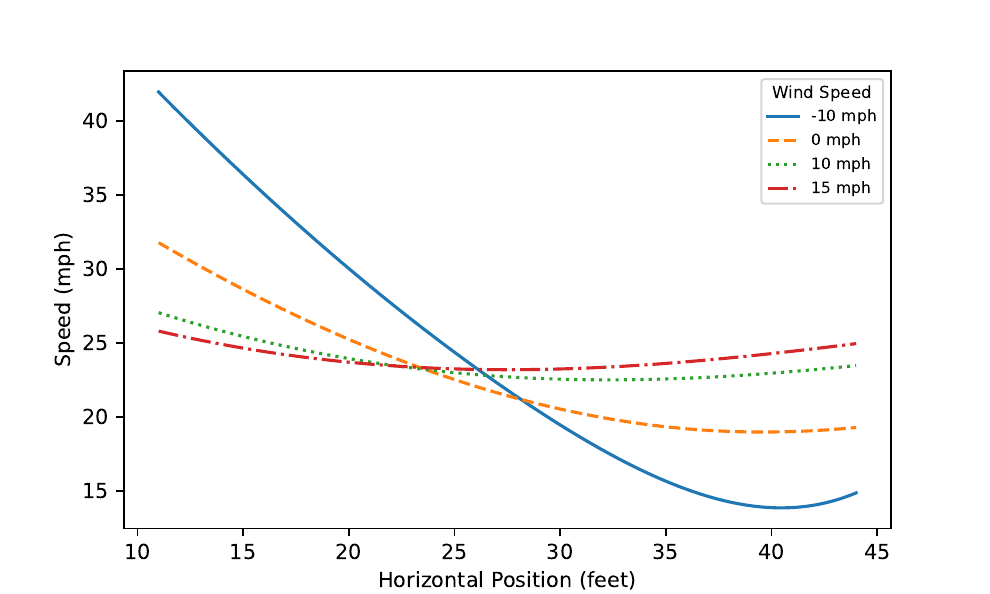}
    \caption{Speed of an optimally hit drive shot in four wind
    conditions, $w=-10$ mph, $w=0$ mph, $w=10$ mph and $w=15$ mph plotted against
    the horizontal location.
    Other input values are set at
    $x_0 = 11$ feet, $y_0=3$ feet and $\theta = 20$ degrees.}
    \label{fig:speed}
\end{figure}

We now consider an exercise with the same input values as given
in Section~\ref{subsec:traj}. However, this time we calculate the velocity
functions
$x'(t, \theta, w, v_0)$
and 
$y'(t, \theta, w, v_0)$.  
We evaluate the coordinate velocities $x'$ and $y'$
for increasing times $t$. Then, the overall
speed $v$ is calculated via $v = [ (x')^2 + (y')^2]^{1/2}$.
In Figure~\ref{fig:speed}, we plot $v$ versus the horizontal location $x$
under the wind conditions
$w=-10$ mph, $w=0$ mph, $w=10$ mph and $w=15$ mph.

In Figure~\ref{fig:speed}, we again observe that the condition of
playing against wind (i.e.\ $w=-10$ mph) is significantly
different from the other three cases.
For example, the initial velocity $v_0$ is greatest when playing
against the wind. This is necessary in order for the shot to
reach the right endline.
When playing against the wind, we also observe greater initial
deceleration (i.e.\ the slope of the velocity curve is steeper).
On the other hand,
when playing with the wind, the ball maintains a similar velocity
throughout its path.
Under all four wind conditions, we see that the pickleball
speed is similar (approximately 24 mph) 
at the horizontal position 27 feet. The 27-foot position
is close to the boundary
of the non-volley zone (NVZ) on the right hand side of the court.
From a playing perspective, this is interesting since the NVZ boundary
is widely regarded as being the most strategic
position.

\subsection{Strategy - Playing in the Wind}
\label{subsec:wind}

Here we return to the primary strategic question: 
is it better to play with the
wind or against the wind?

As mentioned at the beginning of Section~\ref{subsec:simple}, there are various shots in pickleball. Some shots are infrequent
(e.g.\ lob shots). Smash shots are also less common than
other shots, although it is apparent that a trailing wind makes
smash shots even faster (i.e.\ more difficult to handle).
Alternatively, some shots are not greatly affected by the wind. For example, dink
shots are soft shots taken close to the net; consequently, 
they are not in the air
for long periods of time. A case could be made that it is
preferable to play against the wind when hitting the common drop shot.
Against the wind, a player needs to
worry less about ``popping up'' their drop shot and having it smashed back.
The drop shot will be pushed down by the wind.
Therefore, before endorsing playing against the
wind over playing with the wind, we need to look at the common
drive shot.

We now consider
the merits of playing against the wind versus playing with the
wind in the context of drive shots.
For drive shots, we assume that the player of interest has played optimally in the
sense that the ball is hit hard enough to bounce on the right endline
should it be left untouched.

We use the general approach described in Section~\ref{subsec:simple} to evaluate
the time that it takes the ball to reach the opponent (i.e.\ point
of interest in Figure~\ref{fig:Setup}). 
If it takes less time to reach the opponent
playing with the wind, then playing with the wind is preferred. 
If it takes less time to reach the opponent
playing against the wind, then playing against the wind is preferred. 
We calculate time differences under the following conditions:
$x_0 = 0, 11, 15$ feet (corresponding to endline, mid-court and non-volley
zone) for the player executing the shot,
$z_0 = 29, 33, 44$ feet (corresponding to non-volley zone, mid-court
and endline) for the opponent,
launch angle $\theta = 20$ degrees and launch height $y_0=3$ feet.

Letting $t_w$ denote the time in seconds that it takes the ball to reach
the opponent with an assisting wind $w \geq 0$, we
consider the excess time difference 
$D_w = t_w -t_{-w}$ that it takes for the ball to
reach the opponent when playing with the wind compared to 
when playing against the wind.
This is evaluated for the wind conditions
$w=10$ mph, $w=15$ mph and $w=20$ mph.
Table~\ref{tb:times} provides the results.
We note that the time
difference results in Table~\ref{tb:times} are not greatly sensitive
to minor modifications in the values of $\theta$ and $y_0$.

\begin{table}[ht]
    \begin{center}
        \begin{tabular}{r r | r r r}
            $x_0$ & $z_0$ & $D_{10}$ & $D_{15}$ & $D_{20}$ \\ \hline
            0  & 29 &  0.097 &  0.200 &  0.346 \\
            0  & 33 &  0.056 &  0.166 &  0.348 \\
            0  & 44 & -0.302 & -0.490 &  0.089 \\
            11 & 29 &  0.082 &  0.150 &  0.242 \\
            11 & 33 &  0.058 &  0.133 &  0.255 \\
            11 & 44 & -0.230 & -0.375 & -0.568 \\
            15 & 29 &  0.075 &  0.131 &  0.203 \\
            15 & 33 &  0.059 &  0.124 &  0.223 \\
            15 & 44 & -0.203 & -0.332 & -0.505 \\
        \end{tabular}
        \caption{Excess time difference in seconds $D_w$ that it
        takes the drive shot to reach the opponent when playing with the wind compared
        to when playing against the wind where $w$ is recorded in mph.
        The calculations are carried out using 9 combinations of
        $x_0$ and $z_0$, and using typical settings
        $\theta = 20$ degrees and $y_0= 3$ feet.}
        \label{tb:times}
    \end{center}
\end{table}

From Table~\ref{tb:times}, we observe that most of the $D_w$ entries are positive. This
suggests that there is a competitive advantage to playing against
the wind when hitting the
common drive shot. The ball reaches the opponent faster and there is less
time for the opponent to react when playing against the wind. 
The only situations where $D_w$ is negative correspond to
the setting $z_0 = 44$ feet (i.e.\ the opponent is located on the right endline).
This is noteworthy since it is generally accepted pickleball strategy
to approach the non-volley zone, and not sit back at the right endline.

It is also interesting to look at the row with input settings
$x_0 =15$ feet and $z_0 = 29$ feet. This corresponds to the common
situation where both players have approached the non-volley zone
and are as close as possible. Here, we see that as the wind $w$ increases,
$D_w$ increases. That is, the advantage of playing against the
wind becomes greater as the wind blows harder.
In fact, this same phenomenon is observed in all situations
in Table~\ref{tb:times} whenever $z_0 \neq 44$ feet.

\section{DISCUSSION}
\label{sec:Discussion}

This paper appears to be the first serious investigation of
flight dynamics in the sport of pickleball. Our main contribution is 
one of strategy; we argue that playing against wind is generally
preferable to playing with the wind. Previously, there appeared
to be no consensus opinion on the preference.
The work is based on
a detailed physical model that takes into account relevant
inputs including air resistance and wind. Python code is provided
in a github page (see the Appendix) that allows researchers
to graph pickleball trajectories and velocities under various conditions.

Although the results provided in this paper correspond to our
intuition and were derived from existing knowledge of projectile
motion, it would be good to verify some of the results against
video taken from pickleball matches.
In future research,
it may also be useful to consider additional wind environments
such as crosswinds.


\section{REFERENCES}

\begin{description}
\begin{small}

\item[[1]]
``2023 Sports, Fitness, and Leisure Activities Topline Participation Report''.
Sports \& Fitness Industry Association, 2024.

\item[[2]]
Jim Albert, Mark E Glickman, Tim B Swartz, and Ruud Koning.
``Handbook of statistical methods and analyses in sports''. CRc Press, 2017.

\item[[3]]
Peter Chudinov.
``Projectile motion in a medium with quadratic drag at constant horizontal wind''.
2022. arXiv: 2206.02397.

\item[[4]]
John R Dormand and Peter J Prince.
``A family of embedded Runge-Kutta formulae''.
{\it Journal of Computational and Applied Mathematics} 6.1 (1980), pp. 19-26.

\item[[5]]
Paramjit S Gill and Tim B Swartz.
``A characterization of the degree of weak and strong links in doubles sports''.
{\it Journal of Quantitative Analysis in Sports} 15.2 (2019), pp. 155-162.

\item[[6]]
Marko V Lubarda and Vlado A Lubarda.
``A review of the analysis of wind-influenced projectile 
motion in the presence of linear and nonlinear drag force''.
{\it Archive of Applied Mechanics} (2022), pp. 1997-2017.

\item[[7]]
Bruce R Munson, Donald F Young, and Theodore H Okiishi.
``Fundamentals of Fluid Mechanics''.
John Wiley \& Sons Inc, 1997.

\item[[8]]
Jenn Rossmann and Andrew Rau.
``An experimental study of wiffle ball aerodynamics''.
{\it American Journal of Physics - AMER J PHYS} (Dec. 2007).
DOI: 10.1119/1.2787013.

\item[[9]]
Pauli Virtanen et al.
``SciPy 1.0: Fundamental Algorithms for Scientific Computing in Python''.
{\it Nature Methods} 17 (2020), pp. 261-272. 
DOI: 10.1038/s41592-019-0686-2.

\end{small}
\end{description}

\section{APPENDIX}
\label{sec:Appendix}

This section provides details regarding the
pickleball motion model. It is a projectile equation which takes into account the air resistance and wind speed. A similar model has been used in [3] to study the projectile motion in three other sports: badminton, tennis and golf. Before presenting the full mathematical equation, we introduce and recall previous notation related to the pickleball and its motion:
\begin{itemize}
\item $m$: mass 

\item $t$: flight time 

\item $(x, y)$: coordinates 

\item $\vec{v} = (x', y')$: velocity 

\item $v = \abs{\vec{v}}$: speed 

\item $v_0$: initial speed 

\item $\theta$: initial launch angle 

\item $w$: horizontal constant wind speed.

\end{itemize}

\begin{figure}[ht]
    \centering
    \includegraphics[scale=0.65]{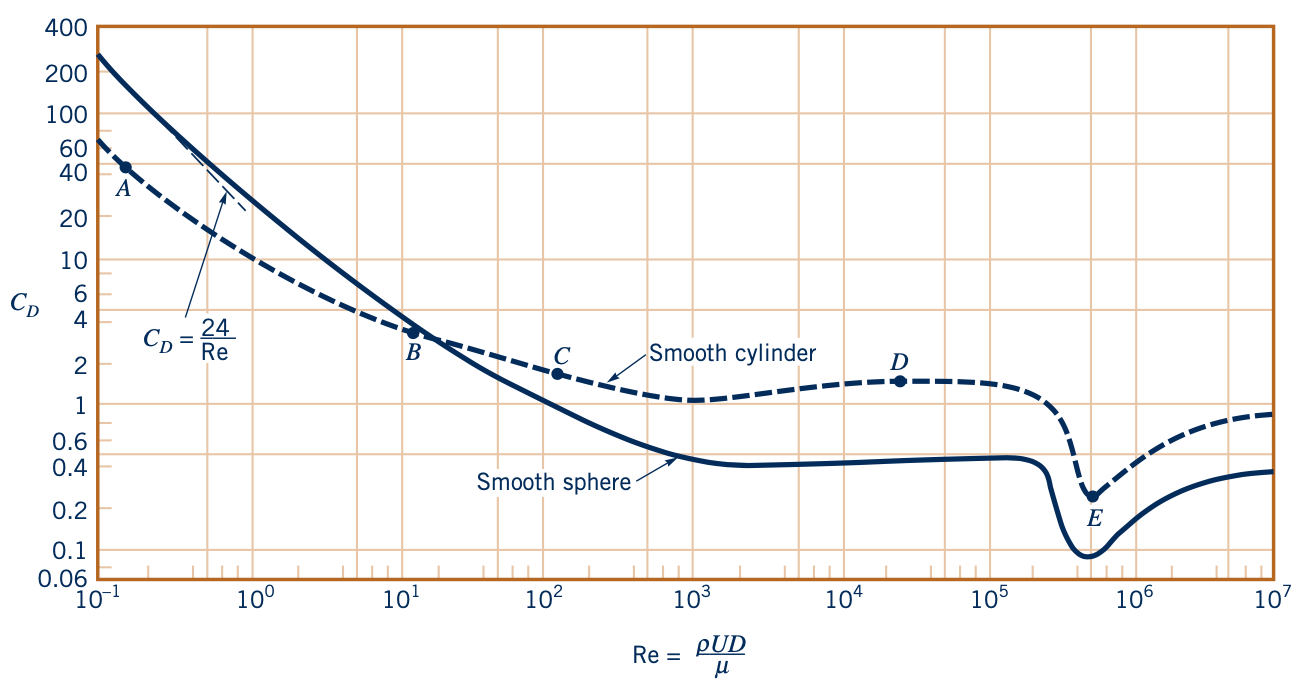}
    \caption{{\em Fundamentals of Fluid Mechanics}, Munson et al, page 520}
    \label{fig:Cd}
\end{figure}

The equation for projectile motion follows from Newton's second law, where we only take into account the gravity and air resistance acting on the pickleball. The air resistance or the drag force is given by 
\begin{align} \label{drag-force}
   \vec{F} = -\frac{1}{2} \rho \, C_d \, A \, (\vec{v} - (w, 0))  \abs{\vec{v} - (w, 0)},
\end{align}
where 
\begin{itemize}
\item $\rho$ is the density of the air, 
\item $A$ is the cross-sectional area of the pickleball, 
\item $\vec{v} - (w, 0)$ is the relative velocity of the pickleball with respect to the wind,
\item $C_d$ is the drag coefficient. 
\end{itemize}
The drag coefficient varies with the Reynolds number 
\begin{align} \label{def:Re}
  Re = \frac{\rho \, U D}{\mu},
\end{align}
where $U$ is the pickleball speed relative to the air, $D$ is the diameter of the ball, $\rho$ is the air density, and $\mu$ is the dynamic viscosity of the atmosphere. The relation between $C_d$ and $Re$ is in general a complicated nonlinear function that depends on the object shape, the object orientation, and characteristics of the air flow. Examples of \(C_d\) for a smooth cylinder and a smooth sphere are shown in Figure~\ref{fig:Cd}, which was taken from Munson et al.\ [7]. 

Though the functional form of \(C_d\) can vary depending on the situation, it is roughly proportional to \(Re^{-1}\) for low Reynolds numbers while turbulence, or irregular air motion, is minimal. At larger Reynolds numbers, when there is significant turbulence, \(C_d\) evens out to stay roughly constant. As we can see from Fig.~\ref{fig:Cd}, the \(C_d\) value for a smooth sphere stays on the same order of magnitude from a Reynolds number of about \(10^3\) onward, though with a dip around \(10^5\) before returning to its constant behaviour. Rougher surfaces tend to lower this threshold Reynolds number by increasing the turbulence around the sphere. Thus, we expect the holes in a pickleball to reduce the Reynolds number required to produce a roughly constant \(C_d\) to a value even lower than \(10^3\). 

The parameters in our problem correspond to $Re \gtrapprox 2.5 \times 10^4$, which is well above the threshhold of $10^3$. Thus we conclude that \(C_d\) should stay roughly constant. In terms of the exact value of this constant, since we could not find any experimental measurements of \(C_d\) for pickleballs, we approximated the \(C_d\) value by treating the pickleball as a forward-facing wiffleball and using the experimental results found by Rossmann et al.\ [8]. This gave us a constant \(C_d\) of approximately 0.6 to use in Equation~(\ref{drag-force}). The constant drag coefficient leads to a quadratic dependence of the drag force on the the relative velocity instead of a linear one as often used in projectile equations. 
The full system then has the form
\begin{align} \label{eq:pickleball}
    \begin{split}
          & m x''(t) = - \frac{1}{2} \rho \, C_d \, A \, (x'(t) - w) \sqrt{(x'(t) - w)^2 + (y'(t))^2}, 
        \\
          & m y''(t) = - g - \frac{1}{2} \rho \, C_d \, A \, (y'(t) - w) \sqrt{(x'(t) - w)^2 + (y'(t))^2},
        \\
          & (x(0), y(0)) = (x_0, y_0), 
        \qquad
          (x'(0), y'(0)) = (v_0 \cos\theta, v_0 \sin\theta),
    \end{split}
\end{align}
where \(C_d\) is 0.6 as stated before, $g$ is the gravitational constant \(9.18~\mathrm{m}/\mathrm s^2\), \(\rho\) is a standard atmospheric density of \(1.2~\mathrm{kg}/\mathrm{m}^3\), \(A\) is a standard cross-sectional area for a pickleball of \(\pi \times (37~\mathrm{mm})^2\), \(m\) is a standard pickleball mass of \(24\) g, $(x', y')$ denote the first order time derivatives which give the velocity and $(x'', y'')$ denote the second order time derivatives which give the acceleration of the ball. 

System~\eqref{eq:pickleball} is solved numerically using the explicit Runge-Kutta method of order 5(4) provided by default in Scipy's [9] \texttt{solve\_ivp} function [4]. The initial speed $v_0$ given implicitly by conditions in~(\ref{eq:simple1})
are determined by using Scipy's \texttt{fsolve} function to numerically solve for the roots of \((x - 44, y)\), using the \(x(t), y(t)\) functions we found. The Python code used to accomplish this is hosted at \url{https://github.com/0Strategist0/Pickleball}. 

It should be noted that though our choice of \(C_d\) is reasonable given the data we had and seems to produce pickleball trajectories and velocities similar to what is often measured, the true \(C_d\) for a pickleball could in principle vary by roughly \(\pm 0.5\) in certain conditions. We did test several such alternate \(C_d\) values, and the exact numerical values for the time differences $D_w$ we obtained could be significantly different than the ones shown in this paper. However, the signs of all these time differences were preserved after varying \(C_d\), meaning that our main conclusions about whether to play with or against the wind seem to hold regardless of the specific value of \(C_d\). It would be interesting for future work to obtain experimental data measuring \(C_d\) at a variety of Reynolds numbers, allowing for comparison with our model and the computation of more accurate numerical results.

\end{document}